\documentclass[preprint,12pt]{elsarticle}
\usepackage{latexsym,amssymb,amsmath,graphicx,epstopdf,epsf,epsfig,bbm,float,subfig}
\usepackage{figNoNum}
\usepackage[printfigures]{figcaps}
\usepackage{epstopdf}

\newcommand{\be}{\begin{equation}}
\newcommand{\ee}{\end{equation}}
\newcommand{\bea}{\begin{eqnarray}}
\newcommand{\eea}{\end{eqnarray}}
\newcommand{\nn}{\nonumber}

\newcommand{\bax}{\overline{x_t}}
\newcommand{\baxs}{\overline{x^2_t}}
\newcommand{\tih}{\tilde{h}}
\newcommand{\deh}{\delta h}
\newcommand{\eps}{\epsilon}

\renewcommand{\vec}[1]{\mbox{\boldmath${#1}$}}

\newcommand{\ei}{\end{itemize}}
\newcommand{\bi}{\begin{itemize}}

\newcommand{\vu}{\mbox{$\vec{u}$}}

\newcommand{\MB}{\left[\begin{array}}
\newcommand{\ME}{\end{array}\right]}

\newcommand{\defi}{\stackrel{\bigtriangleup}{=}}
\makeatletter

\newcommand{\Rmnum}[1]{\expandafter\@slowromancap\romannumeral #1@}
\makeatother

\journal{Digital Signal Processing}

\begin{document}

\begin{frontmatter}
\title{Robust Estimation in Rayleigh Fading Channels Under Bounded Channel Uncertainties}
\author[1]{Mehmet A. Donmez\corref{cor1}}
\cortext[cor1]{Corresponding Author.}
\ead{medonmez@ku.edu.tr}
\author[1]{Huseyin A. Inan}
\ead{hinan@ku.edu.tr}
\author[1]{Suleyman S. Kozat}
\ead{skozat@ku.edu.tr}
\address[1]{Department of ECE, Koc University, Istanbul, Tel: 90 212 3381864}

\begin{abstract}
We investigate channel equalization for Rayleigh fading channels under
bounded channel uncertainties. We analyze three robust methods to estimate an unknown signal transmitted
through a Rayleigh fading channel, where we avoid directly tuning the equalizer parameters to the available
inaccurate channel information. These methods are based on minimizing certain mean-square error
criteria that incorporate the channel uncertainties into the problem formulations.
We present closed-form solutions to the channel equalization problems for each
method and for both zero mean and nonzero mean signals. We illustrate the performances of the
equalization methods through simulations.
\end{abstract}
\end{frontmatter}
\section{Introduction}\label{sec:intro}
The problem of estimating an unknown desired signal transmitted through
an unknown linear time-invariant channel is heavily investigated in the signal processing literature
\cite{dsp1,dsp2,yonina2,yonina3,pilanci,KoEr10}.
Although the underlying channel impulse response is not known exactly in general, an estimate and
an uncertainty bound on it are given \cite{yonina2,yonina3,pilanci}.
In this paper we study the channel equalization problem for Rayleigh fading channels under
bounded channel uncertainties \cite{proakis} and investigate three robust methods to equalize Rayleigh
fading channels.
We avoid directly tuning the equalizer parameters to the inaccurate channel information that is available.
These channel equalization frameworks we investigate are based on minimizing certain mean-square error (MSE) criteria, which incorporate
the channel uncertainties into the problem formulations.
The first approach we investigate is the affine minimax equalization method \cite{yonina3,yonina4,huber1},
which minimizes the estimation error for the worst case channel perturbation.
The second approach we study is the affine minimin equalization method \cite{pilanci,schubert},
which minimizes the estimation error for the most favorable perturbation.
The third approach is the affine minimax regret equalization method \cite{yonina2,yonina3,nargiz,KoEr10},
which minimizes a certain ``regret'' as defined in Section~\ref{sec:sys} and further detailed in Section~\ref{sec:frames}.
We provide closed-form solutions to the affine minimax equalization, the minimin equalization and the minimax regret equalization problems
for both zero mean and nonzero mean signals. Note that the nonzero mean signals frequently appear in iterative equalization applications \cite{nargiz,singer_turbo} and equalization with these signals under channel uncertainties is particularly important and challenging.

When there are uncertainties in the channel coefficients, one of the prevalent approaches to find a
robust solution to the equalization problem is the minimax equalization method \cite{huber1,yonina3,yonina4}.
In this approach, affine equalizer coefficients are chosen to minimize the MSE with respect to the
worst possible channel in the uncertainty bounds. We emphasize that although the minimax equalization
framework has been introduced in the context of statistical signal processing literature
\cite{huber1,yonina3,yonina4}, our analysis significantly differs since we provide a closed-form solution
to the minimax equalization problem for Rayleigh fading channels. In \cite{yonina3}, the uncertainty is in
the noise covariance matrix and the channel coefficients are assumed to be perfectly known. Furthermore, note
that in \cite{yonina4}, the minimax estimator is formulated as a solution to a semidefinite programming (SDP) problem,
unlike in here. In this paper, the uncertainty is in the channel impulse response and we provide a explicit
solution to the minimax channel equalization problem.

Although the minimax equalization method is able to minimize the estimation error for the worst case channel
perturbation, however, it usually provides unsatisfactory results on the average \cite{pilanci}.
An alternative approach to the channel equalization problem is the minimin equalization method
\cite{pilanci,schubert}. In this approach, equalizer parameters are selected to minimize the MSE with
respect to the most favorable channel over the set of allowed perturbations. Although the minimin approach has been
studied in the literature \cite{pilanci,schubert}, however, we emphasize that to the best of
our knowledge, this is the first closed-form solution to the minimin channel equalization problem for Rayleigh
fading channels.

The minimin approach is highly optimistic, which could yield unsatisfactory results, when the difference between the
underlying channel impulse response and the most favorable channel impulse response is relatively
high \cite{pilanci}. In order to preserve robustness and counterbalance the conservative nature of the minimax approach,
the minimax regret approaches have been introduced in the signal processing literature \cite{yonina2, Ko08, KoEr10}.
In this approach, a relative performance measure, i.e., ``regret", is defined as the difference between the MSE of an affine
equalizer and the MSE of the affine minimum MSE (MMSE) equalizer \cite{KoEr10}. The minimax regret channel equalizer seeks
an equalizer that minimizes this regret with respect to the worst possible channel in the uncertainty region.
Although this approach has been investigated before, the minimax regret estimator is formulated as a solution to
an SDP problem \cite{yonina2}, unlike here. In this paper, we explicitly provide the equalizer coefficients and the estimate of the desired
signal.

Our main contributions are as follows. We first formulate the affine equalization problem
for Rayleigh fading channels under bounded channel uncertainties. We then investigate three robust
approaches; affine minimax equalization, affine minimin equalization, and affine minimax regret equalization
for both zero mean and non-zero mean signals.
The equalizer coefficients, and hence, the MSE of each methods have been explicitly provided,
unlike in \cite{yonina2, yonina3, yonina4, pilanci, KoEr10}.

The paper is organized as follows. In Section \ref{sec:sys}, the basic transmission system is
described, along with the notation used in this paper. We present the affine
equalization approaches in Section \ref{sec:frames}. First, we study the affine minimax equalization
tuned to the worst possible channel filter. We then investigate the minimin approach and the minimax
regret approach, and provide the explicit solutions to the corresponding optimization problems. In addition, we present and compare
the MSE performances of all robust affine equalization methods in Section \ref{sec:sim}.
Finally, we conclude the paper with certain remarks in Section \ref{sec:conc}.

\section{System Description}\label{sec:sys}
In this section, we provide the basic description of the system studied in this paper. Here, the signal $x_t$ is transmitted through a discrete-time time-varying channel with an impulse response $h_t$. The received signal $y_t$ is given by
\begin{equation}
y_t = x_t h_t + n_t,
\label{eq:filter}
\end{equation}
where the observation noise $n_t$ is independent and identically distributed (i.i.d.) with zero mean and variance $\sigma_n^2$ and independent from $x_t$. However, instead of the true channel impulse response, an estimate of $h_t$ is provided as $\tilde{h}_t$, where $\deh_t\defi \tih_t - h_t$ is the uncertainty in the channel impulse response and is modeled by $\vert h_t - \tih_t \vert=\vert \deh_t \vert\leq\epsilon$, $\epsilon > 0$, $\epsilon < \infty$, where $\epsilon$ or a bound on $\epsilon$ is known.
\begin{figure}
\centering
\includegraphics[scale=0.6]{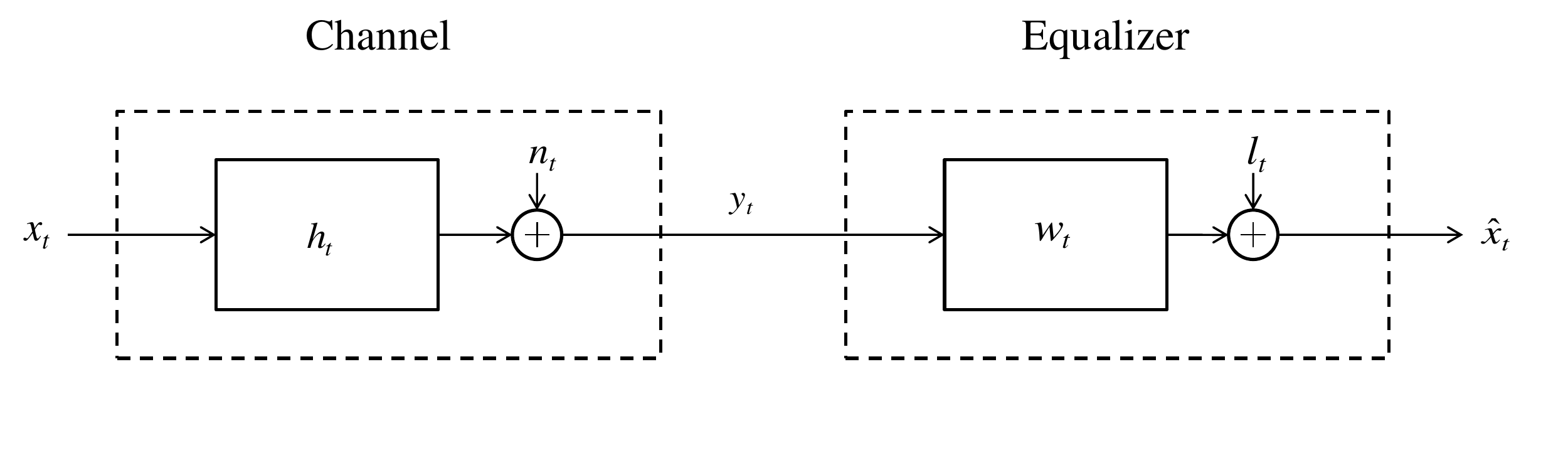}
\caption{A Basic Affine Equalizer Framework}
\end{figure}

We then use the received signal $y_t$ to estimate the transmitted signal $x_t$ as shown in Fig.~1. The estimate of the desired signal is given by
\begin{align}
\hat{x}_t &= w_t y_t + l_t \nn \\
          &= w_t(x_t h_t + n_t)+l_t\label{eq:equalizer},
\end{align}
where $w_t$ is the equalizer impulse response. We note that in \eqref{eq:equalizer}, the equalizer is ``affine" where there is a bias term $l_t$ since the transmitted signal $x_t$, and consequently the received signal $y_t$, are not necessarily zero mean and the mean sequence ${\bar{y}_t}\defi E[y_t]$ is not known due to uncertainty in the channel.

Even under the channel uncertainties, the equalizer impulse response $w_t$ and the bias term $l_t$ can be simply optimized to minimize the MSE for the channel that is tuned to the estimate $\tih_t$, which is also known as the MMSE estimator \cite{linest}. The corresponding impulse response and the bias term are given by
\begin{align*}
\{w_{0,t},l_{0,t}\} = \arg \min_{w,l} E\left[ \left(x_t - w\left(\tih_t x_t + n_t\right) - l\right)^2 \right].
\end{align*}
However, the estimate
\begin{align*}
\hat{x}_{0,t} \defi w_{0,t} y_t + l_{0,t}
\end{align*}
may not perform well when the error in the estimate of the channel impulse response is relatively high \cite{ghaoi,yonina2,yonina3}. One alternative approach to find a robust solution to this problem is to minimize a worst case MSE, which is known as the minimax criterion, as
\begin{align*}
\{w_{1,t},l_{1,t}\} = \arg \min_{w,l} \max_{\vert\deh_t\vert\leq\eps} E\left[ \left(x_t - w\left((\tih_t+\deh_t) x_t + n_t\right) - l\right)^2 \right],
\end{align*}
where $w_{1,t}$ and $l_{1,t}$ minimize the worst case error in the uncertainty region. However, this approach may yield highly conservative results, since the estimate
\begin{align*}
\hat{x}_{1,t} \defi w_{1,t} y_t + l_{1,t}
\end{align*}
is formed by using the equalizer impulse response $w_{1,t}$ and the bias term $l_{1,t}$ that minimize the worst case error, i.e., the error under the worst possible channel impulse response \cite{pilanci,yonina2,yonina3}. Instead of this conservative approach, another useful method to estimate the desired signal is the minimin approach, where the equalizer impulse response and the bias term are given by
\begin{align*}
\{w_{2,t},l_{2,t}\} = \arg \min_{w,l} \min_{\vert\deh_t\vert\leq\eps} E\left[\left(x_t - w\left((\tih_t+\deh_t) x_t + n_t\right) - l\right)^2 \right],
\end{align*}
where $w_{2,t}$ and $l_{2,t}$ minimize the MSE in the most favorable case, i.e., the MSE under the best possible channel impulse response \cite{pilanci}. The estimate of the transmitted signal $x_t$ is given by
\begin{align*}
\hat{x}_{2,t} \defi w_{2,t} y_t + l_{2,t}.
\end{align*}
A major drawback of the minimin approach is that it is a highly optimistic technique, which could yield unsatisfactory results, when the difference between the actual and the best channel impulse responses is relatively high \cite{pilanci}.

In order to reduce the conservative characteristic of the minimax approach as well as to maintain robustness, the minimax regret approach is introduced, which provides a trade-off between performance and robustness \cite{yonina2,nargiz,KoEr10}. In this approach, the equalizer impulse response and the bias term are chosen in order to minimize the worst-case ``regret", where the regret for not using the MMSE is defined as the difference between the MSE of the estimator and the MSE of the MMSE, i.e.,
\begin{align*}
&\{w_{3,t},l_{t,3}\} = \arg \min_{w,l} \max_{\vert\deh_t\vert\leq\eps}\\
&\left\{E\left[ \left(x_t - w\left((\tih_t+\deh_t) x_t + n_t\right) - l\right)^2 \right]- \min_{w,l}E\left[ \left(x_t - w\left((\tih_t+\deh_t) x_t + n_t\right) - l\right)^2 \right] \right\}.
\end{align*}
The corresponding estimate of the desired signal $x_t$ is given by
\begin{align*}
\hat{x}_{3,t} \defi w_{3,t} y_t + l_{3,t}.
\end{align*}

In the next section, we investigate and provide closed form solutions for the three equalization formulations:
\begin{itemize}
\item Affine minimax equalization framework,
\item Affine minimin equalization framework,
\item Affine minimax regret equalization framework.
\end{itemize}
We first solve the corresponding optimization problems and obtain the estimates of the desired signal. We next compare their mean-square error performances in Section~\ref{sec:sim}.

\section{Equalization Frameworks \label{sec:frames}}

\subsection{Affine MMSE Equalization}\label{subsec:MMSE}
In this section, we present the affine MMSE equalization framework for completeness \cite{nargiz,linest}. Since the channel impulse response $h_t$ is not accurately known but estimated by $\tih_t$, a linear equalizer that is matched to the estimate $\tih_t$ and minimizes the MSE can be used to estimate the transmitted signal $x_t$. The corresponding equalizer impulse response $w_{0,t}$ and the bias term $l_{0,t}$ are given by
\begin{align}
\{w_{0,t},l_{0,t}\} = \arg \min_{w,l} E\left[ \left(x_t - w\left(\tih_t x_t + n_t\right) - l\right)^2 \right].
\label{eq:MMSE}
\end{align}
Since the cost function in \eqref{eq:MMSE} is convex in $w$ and $l$, the minimizers $w_{0,t}$ and $l_{0,t}$ are given by
\begin{align*}
w_{0,t} = \frac{\tih_t \sigma_x^2}{\tih_t^2 \sigma_x^2+ \sigma_n^2},\;\;\;l_{0,t} = \frac{\bax \sigma_n^2 }{\tih_t^2 \sigma_x^2+ \sigma_n^2}.
\end{align*}


\subsection{Affine Equalization Using a Minimax Framework}\label{subsec:minimax}
In this section, we investigate a robust estimation framework based on a minimax criteria \cite{linest,chandra1,schubert}. We find the
 equalizer impulse response $w_{1,t}$ and the bias term $l_{1,t}$ that solve the following optimization problem:
\begin{align}
\{w_{1,t},l_{1,t}\} = \arg \min_{w,l} \max_{\vert\deh_t\vert\leq\eps} E\left[ \left(x_t - w\left((\tih_t+\deh_t)
x_t + n_t\right) - l\right)^2 \right].
\label{eq:minmax}
\end{align}
In \eqref{eq:minmax}, we seek to find an equalizer impulse response $w_{1,t}$ and a bias term $l_{1,t}$ that perform best in the worst possible scenario.
This framework can be perceived as a two-player game problem, where one player tries to pick $w_{1,t}$ and $l_{1,t}$ pair that minimize the MSE for a given channel uncertainty while the opponent pick $\deh_t$ to maximize MSE for this pair. In this sense, this problem is constrained since there is a limit on how large the channel uncertainty
$\deh_t$ can be, i.e., $\vert \deh_t \vert\leq\eps$ where $\eps$ or a bound on $\eps$ is known.

In the following theorem we present a closed form solution to the optimization problem \eqref{eq:minmax}.\\
{\bf Theorem 1: } Let $x_t$, $y_t$ and $n_t$ represent the transmitted, received and noise signals such that
 $y_t=h_t x_t +n_t$, where $h_t$ is the unknown channel impulse response and $n_t$ is i.i.d. zero mean with
 variance $\sigma_n^2$. At each time $t$, given an estimate $\tih_t$ of $h_t$ satisfying
 $\vert h_t-\tilde{h}_t\vert\leq\epsilon$, the solution to the optimization problem
\begin{align}
\{w_{1,t},l_{1,t}\} = \arg \min_{w,l} \max_{\vert\deh_t\vert\leq\eps} E\left[ \left(x_t - w\left((\tih_t+\deh_t) x_t + n_t\right) - l\right)^2 \right]
\label{eq:minmaxcostthm}
\end{align}
is given by
\begin{align*}
w_{1,t} = \left\{
     \begin{array}{ll}
       \frac{\left(\tih_t-\eps\right)\sigma_x^2}{\left(\tih_t-\eps\right)^2\sigma_x^2 +\sigma_n^2} & : \tih_t \eps \sigma_x^2<\eps^2 \sigma_x^2+\sigma_n^2\\
       \frac{1}{\tih_t} & : \tih_t \eps \sigma_x^2\geq\eps^2 \sigma_x^2+\sigma_n^2\\
     \end{array}
     \right.
\end{align*}
and
 \begin{align*}
 l_{1,t} = \left\{
     \begin{array}{ll}
      \frac{\bax \sigma_n^2}{\left(\tih_t-\eps\right)^2 \sigma_x^2+\sigma_n^2} & : \tih_t \eps \sigma_x^2<\eps^2 \sigma_x^2+\sigma_n^2\\
       0 & : \tih_t \eps \sigma_x^2\geq\eps^2 \sigma_x^2+\sigma_n^2.\\
     \end{array} \right.
\end{align*}
{\bf Proof: } Here, we find the equalizer impulse response $w_{1,t}$ and the bias term $l_{1,t}$ that solve the optimization problem in \eqref{eq:minmaxcostthm}. To accomplish this, we first solve the inner maximization problem and find the maximizer channel uncertainty $\deh_t^*$. We then substitute $\deh_t^*$ in \eqref{eq:minmaxcostthm} and solve the outer minimization problem to find $w_{1,t}$ and $l_{1,t}$.

We solve the inner maximization problem as follows. We observe that the cost function in \eqref{eq:minmax} can be written as
\begin{align}
&E\left[ \left(x_t - w\left((\tih_t+\deh_t) x_t + n_t\right) - l\right)^2 \right]\label{eq:minmaxcost}\\
&= E\left[x_t^2\right] + w^2\left(\tih_t+\deh_t\right)^2 + l^2-2lE\left[x_t\right] - 2w\left(\tih_t+\deh_t\right)
E\left[x_t^2\right]
+ 2wl\left(\tih_t+\deh_t\right)E\left[x_t\right] \nn\\
&=w^2\left(\tih_t+\deh_t\right)^2 + 2w\left(\tih_t+\deh_t\right)\left(l\bax-\baxs \right) + \mathrm{C_1},\nn
\end{align}
where $\bax\defi E\left[x_t\right]$, $\baxs\defi E\left[x_t^2\right]$ and $\mathrm{C_1}\defi E\left[x_t^2\right]
+ w^2\sigma_n^2+l^2-2l\bax $ does not depend on $\deh_t$. If we define $a=\baxs>0$, $b=l\bax-\baxs$,
$u=w\left(\tih_t+\deh_t\right)$ and $\mathrm{C_2}=\mathrm{C_1}-\frac{b^2}{a}$, then \eqref{eq:minmaxcost}
can be written as
\begin{align*}
E\left[ \left(x_t - w\left((\tih_t+\deh_t) x_t + n_t\right) - l\right)^2 \right] = a\left(u+\frac{b}{a}\right)^2
+\mathrm{C_2},
\end{align*}
where $\mathrm{C_2}$ is independent of $\deh_t$. Hence the inner maximization problem in \eqref{eq:minmax} can
be written as
\begin{align}
\deh_t^*=\arg \max_{\vert\deh_t\vert\leq\eps} E\left[ \left(x_t - w\left((\tih_t+\deh_t) x_t + n_t\right)
- l\right)^2 \right]
= \arg \max_{\vert\deh_t\vert\leq\eps} \left|w\right|\left|\deh_t + \frac{l\bax-\baxs}{w\baxs} \right|. \label{eq:minmaxcost2}
\end{align}
If we apply the triangular inequality to the second term in \eqref{eq:minmaxcost2}, then we get the following upper bound:
\begin{align*}
\left|w\right|\left|\deh_t + \tih_t + \frac{l\bax-\baxs}{w\baxs} \right|
\leq \left|w\right|\left[\left|\deh_t\right| + \left|\tih_t+\frac{l\bax-\baxs}{w\baxs}\right|\right]
\leq \left|w\right|\left[\eps + \left|\tih_t+\frac{l\bax-\baxs}{w\baxs}\right|\right],
\end{align*}
where the upper bound is achieved at $\deh_t^*=\eps \mathrm{sgn}\left(\tih_t+\frac{l\bax-\baxs}{w\baxs}\right)$,
 where $\mathrm{sgn}(z)=1$ if $z\geq0$ and $\mathrm{sgn}(z)=-1$ if $z<0$. Hence it follows that
\begin{align}
\deh_t^*
=\arg \max_{\vert\deh_t\vert\leq\eps} E\left[ \left(x_t - w\left((\tih_t+\deh_t) x_t + n_t\right) - l\right)^2 \right]
= \left\{
     \begin{array}{ll}
       \eps & : \tih_t+\dfrac{l\bax-\baxs}{w\baxs}\geq0,\\
      -\eps & : \tih_t+\dfrac{l\bax-\baxs}{w\baxs}<0.\\
     \end{array}
     \right. \label{eq:minmaxdh}
\end{align}

We next solve the outer minimization problem as follows. We first note that the minimum in \eqref{eq:minmaxcostthm} is taken over all $w\in\mathbb{R}$ and $l\in\mathbb{R}$. If we write $\vu = \left[w,l\right]^T\in\mathbb{R}^2$ in a vector form and define $\mathcal{U} = \left\{\vu=\left[w,l\right]^T\in\mathbb{R}^2\;\vert\;\tih_t+\frac{l\bax-\baxs}{w\baxs}\geq0\right\}$, then it follows that $\mathcal{V} \defi \left\{\vu=\left[w,l\right]^T\in\mathbb{R}^2\;\vert\;\tih_t+\frac{l\bax-\baxs}{w\baxs}<0\right\}=\mathbb{R}^2 \setminus\mathcal{U}$, i.e., $\mathcal{U} \cup \mathcal{V}=\mathbb{R}^2$ and $\mathcal{U} \cap \mathcal{V}= \varnothing$. Hence, the cost function in the outer minimization problem in \eqref{eq:minmaxcostthm} is given by
\begin{align*}
&\max_{\vert\deh_t\vert\leq\eps} E\left[ \left(x_t - w\left((\tih_t+\deh_t) x_t + n_t\right) - l\right)^2 \right]\\
&=\left\{
     \begin{array}{ll}
       E\left[ \left(x_t - w\left((\tih_t+\eps) x_t + n_t\right) - l\right)^2 \right] & : \left[w,l\right]^T\in\mathcal{U},\\
       E\left[ \left(x_t - w\left((\tih_t-\eps) x_t + n_t\right) - l\right)^2 \right] & : \left[w,l\right]^T\in\mathcal{V}\\
     \end{array}
     \right..
\end{align*}
We first substitute $\deh_t = \eps$ and find the corresponding $\{w,l\}$ pair
that minimizes the objective function in \eqref{eq:minmaxcostthm} to check whether $\left[w,l\right]\in\mathcal{U}$.
We next substitute $\deh_t = -\eps$ and find the corresponding $\{w,l\}$ to check whether
$\left[w,l\right]\in\mathcal{V}$. Based on these criteria, we obtain the corresponding equalizer
impulse response and the bias term pair $\{w_{1,t},l_{1,t}\}$.

We first substitute $\deh_t = \eps$ in the objective function of \eqref{eq:minmaxcostthm} to get the following
minimization problem:
\begin{align}
&\{w^*,l^*\} =\arg \min_{w,l}\\
&\left\{\baxs + w^2 \left( \left(\tih_t + \eps\right)^2 \baxs + \sigma_n^2 \right) + l^2 - 2l \bax + 2wl \left(\tih_t + \eps\right) \bax  - 2w \left(\tih_t + \eps\right)  \baxs\right\}. \label{eq:minmaxconvex1}
\end{align}
We observe that the cost function in \eqref{eq:minmaxconvex1} is a convex function of $w$ and $l$ yielding
\begin{align*}
w^*&=\frac{\left(\tih_t +  \eps\right)\sigma_x^2 }{\left(\tih_t + \eps\right)^2 \sigma_x^2 +  \sigma_n^2},\;\;
l^*=\frac{\bax \sigma_n^2}{\left(\tih_t + \eps\right)^2 \sigma_x^2 +  \sigma_n^2},
\end{align*}
where $\sigma_x^2 \defi E\left[\left(x_t-\bax \right)^2\right]$. However we have
\begin{align}
\frac{ \baxs - l^* \bax}{w^*  \baxs} = \tih_t + \eps + \frac{\sigma_n^2}{(\tih_t +  \eps)\sigma_x^2 } > \tih_t
\end{align}
so that $\left[w^*,l^*\right]^T \not \in \mathcal{U}$.

We next substitute $\deh_t = -\eps$ in the cost function of \eqref{eq:minmaxcostthm} to get
\begin{align}
&\{w^*,l^*\} = \arg \min_{w,l}\\
&\left\{  \baxs + w^2 \left( \left(\tih_t - \eps\right)^2 \baxs + \sigma_n^2 \right) + l^2 - 2l \bax + 2wl \left(\tih_t - \eps\right) \bax  - 2w \left(\tih_t - \eps\right)  \baxs\right\}. \label{eq:minmaxconvex2}
\end{align}
The cost function in \eqref{eq:minmaxconvex2} is also a convex function of $w$ and $l$ so that we get
\begin{align*}
w^*&=\frac{\left(\tih_t -  \eps\right)\sigma_x^2 }{\left(\tih_t - \eps\right)^2 \sigma_x^2 +  \sigma_n^2},\;\;
l^*=\frac{\bax \sigma_n^2}{\left(\tih_t - \eps\right)^2 \sigma_x^2 +  \sigma_n^2}.
\end{align*}
If the condition $\tih_t \eps \sigma_x^2 < \eps^2 \sigma_x^2 + \sigma_n^2$ holds, then we have
\begin{align*}
\tih_t & < \tih_t - \eps  + \frac{\sigma_n^2}{(\tih_t - \eps) \baxs}  < \frac{\baxs - l \bax}{w  \baxs}
\end{align*}
so that $\left[w^*,l^*\right]^T \in \mathcal{V}$. Thus, the corresponding equalizer
impulse response and the bias term are given by
$w_{1,t} = \frac{\left(\tih_t - \eps\right) \sigma_x^2}{\left(\tih_t - \eps\right)^2 \sigma_x^2 + \sigma_n^2}$
and $l_{1,t} = \frac{\bax \sigma_n^2}{(\tih_t - \eps)^2 \sigma_x^2 + \sigma_n^2}$, respectively. However, if
the condition $\tih_t \eps \sigma_x^2 < \eps^2 \sigma_x^2 + \sigma_n^2$ does not hold, then $\left[w^*,l^*\right]^T \not \in \mathcal{V}$ so that the corresponding equalizer impulse response and the bias term are given by are $w_{1,t}= \frac{1}{\tih_t}$ and
$l_{1,t} = \frac{\bax \sigma_n^2}{(\tih_t - \eps)^2 \sigma_x^2 + \sigma_n^2}$. Hence, we have
\begin{align*}
&w_{1,t} = \left\{
     \begin{array}{ll}
       \frac{\left(\tih_t-\eps\right)\sigma_x^2}{\left(\tih_t-\eps\right)^2\sigma_x^2 +\sigma_n^2} & : \tih_t \eps \sigma_x^2<\eps^2 \sigma_x^2+\sigma_n^2\\
       \frac{1}{\tih_t} & : \tih_t \eps \sigma_x^2\geq\eps^2 \sigma_x^2+\sigma_n^2\\
     \end{array}
     \right.,\;\;\;\\
     &l_{1,t} = \left\{
     \begin{array}{ll}
      \frac{\bax \sigma_n^2}{\left(\tih_t-\eps\right)^2 \sigma_x^2+\sigma_n^2} & : \tih_t \eps \sigma_x^2<\eps^2 \sigma_x^2+\sigma_n^2\\
       0 & : \tih_t \eps \sigma_x^2\geq\eps^2 \sigma_x^2+\sigma_n^2\\
     \end{array} \right..
\end{align*}
Hence, the proof follows. $\blacksquare$

In the following corollary, we provide a special case of Theorem 1, where the desired signal $x_t$ is zero mean.\\
{\bf Corollary 1: } When the transmitted signal $x_t$ is zero mean, the solution to the optimization problem
\begin{align}
\{w_{1,t},l_{1,t}\} = \arg \min_{w,l} \max_{\vert\deh_t\vert\leq\eps} E\left[ \left(x_t - w\left((\tih_t+\deh_t) x_t + n_t\right) - l\right)^2 \right]
\label{eq:minmaxcostcor}
\end{align}
is given by
\begin{align*}
w_{1,t} = \left\{
     \begin{array}{ll}
       \frac{\left(\tih_t-\eps\right)}{\left(\tih_t-\eps\right)^2 +\frac{1}{S}} & : \eps\left(\tih_t-\eps\right)<\frac{1}{S}\\
       \frac{1}{\tih_t} & : \eps\left(\tih_t-\eps\right)\geq\frac{1}{S}\\
     \end{array}
     \right.,\;\;\;l_{1,t}=0,
\end{align*}
where $S\defi\sigma_x^2/\sigma_n^2$ is the signal-to-noise ratio (SNR).\\
{\bf Proof: } The proof directly follows from Theorem 1, therefore, is omitted. $\blacksquare$


\subsection{Affine Equalization Using a Minimin Framework}\label{subsec:minimin}

In this section, we study the minimin equalization framework, where the inner maximization of the minimax framework is replaced with a minimization over the uncertainty set \cite{pilanci,chandra2,schubert}. We seek to solve the following optimization problem:
\begin{align}
\{w_{2,t},l_{2,t}\} & = \arg \min_{w,l} \min_{\vert\deh_t\vert\leq\eps} E\left[\left(x_t - w\left((\tih_t+\deh_t) x_t + n_t\right) - l\right)^2 \right].
\label{eq:minmin}
\end{align}

The following lemma is introduced to demonstrate that $\min$ operators in \eqref{eq:minmin} can be interchanged, which will be used in Theorem 2.\\
{\bf Lemma 1:} For an arbitrary function $f(x,y,z)$ and nonempty sets $\mathcal{X}$, $\mathcal{Y}$ and $\mathcal{Z}$, we have
\begin{equation*}
\min_{x\in\mathcal{X},y\in\mathcal{Y}} \min_{z\in\mathcal{Z}} f(x,y,z) = \min_{z\in\mathcal{Z}} \min_{x\in\mathcal{X},y\in\mathcal{X}} f(x,y,z),
\end{equation*}
assuming that all minimums are achieved on the corresponding sets.\\
{\bf Proof: } The proof is straightforward hence omitted here.

In the following theorem we present a closed form solution to the optimization problem \eqref{eq:minmin}.\\
{\bf Theorem 2: } Let $x_t$, $y_t$ and $n_t$ represent the transmitted, received and noise signals such that
 $y_t=h_t x_t +n_t$, where $h_t$ is the unknown channel impulse response and $n_t$ is i.i.d. zero mean with
 variance $\sigma_n^2$. At each time $t$, given an estimate $\tih_t$ of $h_t$ satisfying
 $\vert h_t-\tilde{h}_t\vert\leq\epsilon$, the solution to the optimization problem
\begin{align}
\{w_{2,t},l_{2,t}\} = \arg \min_{w,l} \min_{\vert\deh_t\vert\leq\eps} E\left[ \left(x_t - w\left((\tih_t+\deh_t) x_t + n_t\right) - l\right)^2 \right]
\label{eq:minmincostthm}
\end{align}
is given by
\begin{align*}
w_{2,t} = \frac{\left(\tih_t+\eps \mathrm{sign}(\tih_t)\right)\sigma_x^2}{\left(\tih_t+\eps \mathrm{sign}(\tih_t)\right)^2 \sigma_x^2 +\sigma_n^2}
\end{align*}
and
\begin{align*}
l_{2,t}=\frac{\bax \sigma_n^2}{\left(\tih_t+\eps \mathrm{sign}(\tih_t)\right)\sigma_x^2+\sigma_n^2}.
\end{align*}
{\bf Proof: } Here, we find the equalizer impulse response $w_{2,t}$ and the bias term $l_{2,t}$ that solve the optimization problem in \eqref{eq:minmincostthm}. We first note that, by Lemma 1, we can interchange $\min$ operators in \eqref{eq:minmincostthm} so that the optimization problem in \eqref{eq:minmincostthm} is equivalent to
\begin{align}
&\min_{w,l} \min_{\vert\deh_t\vert\leq\eps} E\left[\left(x_t - w\left((\tih_t+\deh_t) x_t + n_t\right) - l\right)^2 \right]\nn\\
&=\min_{\vert\deh_t\vert\leq\eps} \min_{w,l} E\left[\left(x_t - w\left((\tih_t+\deh_t) x_t + n_t\right) - l\right)^2 \right].\label{eq:minminchange}
\end{align} Hence, we first solve the inner minimization problem in \eqref{eq:minminchange} and find the minimizers $w^*$ and $l^*$. We then substitute $w^*$ and $l^*$ in \eqref{eq:minminchange} and solve the outer minimization problem to find the minimizer $\deh_t^*$, which yields the desired equalizer impulse response $w_{2,t}$ and $l_{2,t}$.

We observe that the objective function in \eqref{eq:minmin} can be written as
\begin{align}
&E\left[ \left(x_t - w\left((\tih_t+\deh_t) x_t + n_t\right) - l\right)^2 \right]\label{eq:minmincostthm}\\
&= E\left[x_t^2\right] + w^2\left(\tih_t+\deh_t\right)^2 + l^2-2lE\left[x_t\right] - 2w\left(\tih_t+\deh_t\right)
E\left[x_t^2\right]\nn\\
&+ 2wl\left(\tih_t+\deh_t\right)E\left[x_t\right] \nn\\
&=\baxs + w^2\left(\left(\tih_t+\deh_t\right)^2 \baxs + \sigma_n^2\right) + \l^2 -2l\bax +2wl\left(\tih_t+\deh_t\right)\bax \nn\\
&- 2w\left(\tih_t+\deh_t\right)\baxs,\nn
\end{align}

We first solve the inner minimization problem in the right hand side of \eqref{eq:minminchange} with respect to $w$ and $l$ as follows. If we define $F(w,l,\deh_t)$\\ $= E\left[ \left(x_t - w\left((\tih_t+\deh_t) x_t + n_t\right) - l\right)^2 \right]$, then we set the first derivatives of $F(w,l,\deh_t)$ with respect to $w$ and $l$ which yields the minimizers $w^*$ and $l^*$, respectively, i.e., $w^*$ and $l^*$ satisfy $\frac{\partial F}{\partial w} \Big|_{w=w^*}=0$ and $\frac{\partial F}{\partial l} \Big|_{l=l^*}=0$. The corresponding partial derivative of the cost function $F(w,l,\deh_t)$ with respect to $l$ is given by
\begin{align*}
\frac{\partial F}{\partial l} \Big|_{l=l^*} = 2l^* - 2\bax + 2w^* \left(\tih_t+\deh_t\right)\bax = 0
\end{align*}
so that $l^*=\bax - w^* \left(\tih_t+\deh_t\right)\bax$. The corresponding partial derivative of $F(w,l,\deh_t)$ with respect to $w$ is given by
\begin{align*}
\frac{\partial F}{\partial w} \Big|_{w=w^*} = 2w^* \left(\left(\tih_t+\deh_t\right)^2\baxs +\sigma_n^2\right) +2l^*\left(\tih_t+\deh_t\right)\bax - 2\left(\tih_t+\deh_t\right)\baxs  = 0,
\end{align*}
which implies that $w^* = \frac{\left(\tih_t+\deh_t\right)\baxs - l^* \left(\tih_t+\deh_t\right)\bax}{\left(\tih_t+\deh_t\right)^2\baxs+\sigma_n^2}$. Thus, we get that
\begin{align*}
w^* &= \frac{\left(\tih_t+\deh_t\right)\sigma_x^2}{\left(\tih_t+\deh_t\right)^2\sigma_x^2+\sigma_n^2}\\
l^* &= \frac{\bax \sigma_n^2}{\left(\tih_t+\deh_t\right)^2\sigma_x^2+\sigma_n^2}
\end{align*}
for a given $\deh_t$.

We next solve the outer minimization problem. If we substitute $w^*$ and $l^*$ in $F(w,l,\deh_t)$, then we obtain
\begin{align}
\deh_t^*
&= \arg \min_{\vert \deh_t \vert\leq\eps} \min_{w,l} E\left[ \left(x_t - w\left((\tih_t+\deh_t) x_t + n_t\right) - l\right)^2 \right] \nn\\
&= \arg \min_{\vert \deh_t \vert\leq\eps} F(w^*,l^*,\deh_t) \nn\\
&= \arg \min_{\vert \deh_t \vert\leq\eps} \frac{\sigma_n^2 \sigma_x^2}{\left(\tih_t+\deh_t\right)^2\sigma_x^2 + \sigma_n^2} \nn \\
&= \arg \max_{\vert \deh_t \vert\leq\eps} \left| \tih_t+\deh_t\right|
\end{align}
so that $\deh_t^* = \eps \mathrm{sign}(\tih_t)$. Hence, the equalizer impulse response $w_{2,t}$ and the bias term $l_{2,t}$ are given by
\begin{align*}
w_{2,t} &= \frac{\left(\tih_t + \eps \mathrm{sign}(\tih_t)\right) \sigma_x^2}{\left(\tih_t + \eps \mathrm{sign}(\tih_t)\right)^2 \sigma_x^2 + \sigma_n^2}\\
l_{2,t} &= \frac{\bax \sigma_n^2}{\left(\tih_t + \eps \mathrm{sign}(\tih_t)\right)\sigma_x^2 + \sigma_n^2}.
\end{align*}
Hence, the proof follows. $\blacksquare$

In the following corollary, we provide a special case of Theorem 1, where the desired signal $x_t$ is zero mean.\\
{\bf Corollary 2: } When the transmitted signal $x_t$ is zero mean, the solution to the optimization problem
\begin{align*}
\{w_{2,t},l_{2,t}\} = \arg \min_{w,l} \min_{\vert\deh_t\vert\leq\eps} E\left[ \left(x_t - w\left((\tih_t+\deh_t) x_t + n_t\right) - l\right)^2 \right]
\end{align*}
is given by
\begin{align*}
w_{2,t} = \frac{\left(\tih_t+\eps \mathrm{sign}(\tih_t)\right)}{\left(\tih_t+\eps \mathrm{sign}(\tih_t)\right)^2 +\frac{1}{S}}
\end{align*}
and
\begin{align*}
l_{2,t}=0,
\end{align*}
where $S\defi \sigma_x^2/\sigma_n^2$ is the SNR.\\
{\bf Proof: }The proof follows from Theorem 2 when $\bax=0$. $\blacksquare$
\subsection{Affine Equalization Using a Minimax Regret Framework}
In this section, we investigate the minimax regret equalization framework, where the performance of an affine equalizer is
defined with respect to the MMSE affine equalizer that is tuned to the unknown channel \cite{yonina2,KoEr10,nargiz,linest}.
We emphasize that the minimax equalization framework investigated in Section~\ref{subsec:minimax} may produce highly
conservative results since the equalizer impulse response $w$ and the bias term $l$ are optimized to minimize the
worst case MSE \cite{linest}. Moreover, the minimin equalization framework introduced in Section~\ref{subsec:minimin}
is a highly optimistic method where the equalizer parameters are optimized to minimize the MSE that corresponds to the
most favorable channel \cite{pilanci}. Thus, the minimin approach may also yield unsatisfactory results in certain applications,
where the channel estimate is highly erroneous \cite{pilanci}. In this context, the minimax regret equalization framework can be used
to improve the equalization performance while preserving the robustness \cite{yonina2,KoEr10}. In this approach, we find the equalizer impulse
response $w_{3,t}$ and the bias term $l_{3,t}$ that solve the following optimization problem:
\begin{align}
&\{w_{3,t},l_{3,t}\} = \arg \min_{w,l} \max_{\vert\deh_t\vert\leq\eps} \nn \\
&\left\{E\left[ \left(x_t - w\left((\tih_t+\deh_t) x_t + n_t\right) - l\right)^2 \right]- \min_{w,l}E\left[ \left(x_t - w\left((\tih_t+\deh_t) x_t + n_t\right) - l\right)^2 \right] \right\}.
\label{eq:minmaxregret}
\end{align}
We note that from Section~\ref{subsec:minimin}, the solution to the minimization problem in the objective function is given by
\begin{align*}
\min_{w,l}E\left[ \left(x_t - w\left((\tih_t+\deh_t) x_t + n_t\right) - l\right)^2 \right]
=\frac{\sigma_n^2 \sigma_x^2}{\left( \tih_t+\deh_t\right)^2\sigma_x^2+\sigma_n^2}.
\end{align*}
Hence the optimization problem in \eqref{eq:minmaxregret} is equivalent to
\begin{align}
&\arg \min_{w,l} \max_{\vert\deh_t\vert\leq\eps}\nn \\
&\left\{E\left[ \left(x_t - w\left((\tih_t+\deh_t) x_t + n_t\right) - l\right)^2 \right]- \min_{w,l}E\left[ \left(x_t - w\left((\tih_t+\deh_t) x_t + n_t\right) - l\right)^2 \right] \right\}\nn\\
&=\arg \min_{w,l} \max_{\vert\deh_t\vert\leq\eps} \left\{E\left[ \left(x_t - w\left((\tih_t+\deh_t) x_t + n_t\right) - l\right)^2 \right]- \frac{\sigma_n^2 \sigma_x^2}{\left( \tih_t+\deh_t\right)^2+\sigma_n^2} \right\}.\label{eq:minmaxregret2}
\end{align}

We first expand the term
\begin{equation*}
\frac{\sigma_n^2 \sigma_x^2}{\left( \tih_t+\deh_t\right)^2\sigma_x^2+\sigma_n^2}
\end{equation*}
in \eqref{eq:minmaxregret2} around $\deh_t=0$ yielding
\begin{equation*}
\frac{\sigma_n^2 \sigma_x^2}{\left( \tih_t+\deh_t\right)^2+\sigma_n^2}
\approx \frac{\sigma_n^2 \sigma_x^2}{\tih_t^2+\sigma_n^2} - \deh_t\frac{2\tih_t\sigma_n^2 \sigma_x^4}{\left(\tih_t^2\sigma_x^2+\sigma_n^2\right)^2}.
\end{equation*}
Hence, instead of \eqref{eq:minmaxregret}, we solve the following optimization problem:
\begin{align}
&\{w_{3,t},l_{3,t}\} = \arg \min_{w,l} \max_{\vert\deh_t\vert\leq\eps} \nn\\
&\left\{E\left[ \left(x_t - w\left((\tih_t+\deh_t) x_t + n_t\right) - l\right)^2 \right] - \frac{\sigma_n^2 \sigma_x^2}{\tih_t^2+\sigma_n^2} + \deh_t\frac{2\tih_t\sigma_n^2 \sigma_x^4}{\left(\tih_t^2\sigma_x^2+\sigma_n^2\right)^2} \right\}, \label{eq:minmaxregret21}
\end{align}
which provides satisfactory results even under large derivations $\delta h_t$ as shown in the Simulations section.

In the following theorem we present a closed form solution to the optimization problem \eqref{eq:minmaxregret21}.\\
{\bf Theorem 3: } Let $x_t$, $y_t$ and $n_t$ represent the transmitted, received and noise signals such that
 $y_t=h_t x_t +n_t$, where $h_t$ is the unknown channel impulse response and $n_t$ is i.i.d. zero mean with
 variance $\sigma_n^2$. At each time $t$, given an estimate $\tih_t$ of $h_t$ satisfying
 $\vert h_t-\tilde{h}_t\vert\leq\epsilon$, the solution to the optimization problem
\begin{align}
&\{w_{3,t},l_{3,t}\} = \arg \min_{w,l} \max_{\vert\deh_t\vert\leq\eps}\nn\\
&\left\{E\left[ \left(x_t - w\left((\tih_t+\deh_t) x_t + n_t\right) - l\right)^2 \right] - \frac{\sigma_n^2 \sigma_x^2}{\tih_t^2+\sigma_n^2} + \deh_t\frac{2\tih_t\sigma_n^2 \sigma_x^4}{\left(\tih_t^2\sigma_x^2+\sigma_n^2\right)^2} \right\}.
\label{eq:minmaxregret3}
\end{align}
is given by
\begin{align*}
&\left[w_{3,t},l_{3,t}\right]= \left\{
        \begin{array}{lll}
            \left[w_1^*,l_1^*\right] & : f\geq0,\; g\geq0,\\
            \left[w_2^*,l_2^*\right] & : f\leq0,\; g\leq0,\\
            \left[w_3^*,l_3^*\right] & : f\geq0,\; g\leq0,\\
            \left[w_4^*,l_4^*\right] & : f<0,\; g>0,
        \end{array}
\right.
\end{align*}\normalsize
where
\begin{align*}
[w_1^*,l_1^*]&=\left[\dfrac{\left(\tih_t+\eps\right)\sigma_x^2}{\left(\tih_t+\eps\right)^2\sigma_x^2+\sigma_n^2},
            \dfrac{\bax\sigma_n^2}{\left(\tih_t+\eps\right)^2\sigma_x^2+\sigma_n^2}\right],\\
[w_2^*,l_2^*]&=\left[\dfrac{\left(\tih_t-\eps\right)\sigma_x^2}{\left(\tih_t-\eps\right)^2\sigma_x^2+\sigma_n^2},
            \dfrac{\bax\sigma_n^2}{\left(\tih_t-\eps\right)^2\sigma_x^2+\sigma_n^2}\right],\\
[w_3^*,l_3^*]&=\arg \min_{\left[w,l\right]\in\left\{\left[w_1^*,l_1^*\right],\left[w_2^*,l_2^*\right]\right\}}\\
&\left\{\max_{\vert\deh_t\vert\leq\eps}\left\{ E\left[ \left(x_t - w\left((\tih_t+\deh_t) x_t + n_t\right) - l\right)^2 \right] - \frac{\sigma_n^2 \sigma_x^2}{\tih_t^2+\sigma_n^2} + \deh_t\frac{2\tih_t\sigma_n^2 \sigma_x^4}{\left(\tih_t^2\sigma_x^2+\sigma_n^2\right)^2} \right\} \right\},\\
[w_4^*,l_4^*]&=\arg \min_{\left[w,l\right]} \left\{E\left[ \left(x_t - w\left(\tih_t x_t + n_t\right) - l\right)^2 \right] - \frac{\sigma_n^2 \sigma_x^2}{\tih_t^2+\sigma_n^2}\right\},\\
f&\defi -\eps-\dfrac{\bax^2 \sigma_n^2}{\left(\tih_t + \eps\right)^2\sigma_x^2+\sigma_n^2} - \dfrac{\sigma_n^2}{\left(\tih_t + \eps\right)\sigma_x^2}
+ \dfrac{\tih_t\sigma_n^2}{\left(\tih_t + \eps\right)^2}\left(\dfrac{\left(\tih_t + \eps\right)^2\sigma_x^2 + \sigma_n^2}{\tih_t^2\sigma_x^2+\sigma_n^2}\right)^2,\\
g&\defi\eps-\dfrac{\bax^2 \sigma_n^2}{\left(\tih_t -\eps\right)^2\sigma_x^2+\sigma_n^2} - \dfrac{\sigma_n^2}{\left(\tih_t - \eps\right)\sigma_x^2}
+ \dfrac{\tih_t\sigma_n^2}{\left(\tih_t - \eps\right)^2}\left(\dfrac{\left(\tih_t - \eps\right)^2\sigma_x^2 + \sigma_n^2}{\tih_t^2\sigma_x^2+\sigma_n^2}\right)^2.
\end{align*}
{\bf Proof: }We first observe that the objective function in \eqref{eq:minmaxregret3} can be written as
\begin{align}
&E\left[ \left(x_t - w\left(\left(\tih_t+\deh_t\right) x_t + n_t\right) - l\right)^2 \right] - \frac{\sigma_n^2 \sigma_x^2}{\tih_t^2+\sigma_n^2} + \deh_t\frac{2\tih_t\sigma_n^2 \sigma_x^4}{\left(\tih_t^2\sigma_x^2+\sigma_n^2\right)^2}\nn\\
&= \baxs + w^2 \left(\tih_t+\deh_t\right)^2\baxs + w^2\sigma_n^2 + l^2 - 2l\bax - 2w\left(\tih_t+\deh_t\right)\baxs + 2wl\left(\tih_t+\deh_t\right)\bax \nn\\
&- \frac{\sigma_n^2 \sigma_x^2}{\tih_t^2+\sigma_n^2} + \deh_t\frac{2\tih_t\sigma_n^2 \sigma_x^4}{\left(\tih_t^2\sigma_x^2+\sigma_n^2\right)^2}\nn\\
&= w^2 \left(\tih_t+\deh_t\right)^2\baxs + \left(\tih_t+\deh_t\right)\left(2wl\bax - 2w\baxs + \frac{2\tih_t\sigma_n^2 \sigma_x^4}{\left(\tih_t^2\sigma_x^2+\sigma_n^2\right)^2}\right) + \mathrm{D}_1, \label{eq:minmaxregretcost}
\end{align}
where $\mathrm{D}_1\defi \baxs + w^2\sigma_n^2 + l^2 - 2l\bax - \frac{\sigma_n^2 \sigma_x^2}{\tih_t^2+\sigma_n^2} - \tih_t\frac{2\tih_t\sigma_n^2 \sigma_x^4}{\left(\tih_t^2\sigma_x^2+\sigma_n^2\right)^2}$ is independent of $\deh_t$. If we define $a=w^2\baxs\geq0$, $b\defi2wl\bax - 2w\baxs + \frac{2\tih_t\sigma_n^2 \sigma_x^4}{\left(\tih_t^2\sigma_x^2+\sigma_n^2\right)^2}$ and $\mathrm{D}_2 = \mathrm{D}_1 - \frac{b^2}{4a}$, then \eqref{eq:minmaxregretcost} can be written as
\begin{align*}
&E\left[ \left(x_t - w\left(\left(\tih_t+\deh_t\right) x_t + n_t\right) - l\right)^2 \right] - \frac{\sigma_n^2 \sigma_x^2}{\tih_t^2+\sigma_n^2} + \deh_t\frac{2\tih_t\sigma_n^2 \sigma_x^4}{\left(\tih_t^2\sigma_x^2+\sigma_n^2\right)^2}\\
&= a\left(u+\frac{b}{2a}\right)^2 + \mathrm{D}_2,
\end{align*}
where $\mathrm{D}_2$ is independent of $\deh_t$. Hence, the inner maximization problem in \eqref{eq:minmaxregret3} is given by
\begin{align}
&\deh_t^* \nn\\
&= \arg \max_{\vert\deh_t\vert\leq\eps} \left\{E\left[ \left(x_t - w\left((\tih_t+\deh_t) x_t + n_t\right) - l\right)^2 \right] - \frac{\sigma_n^2 \sigma_x^2}{\tih_t^2+\sigma_n^2} + \deh_t\frac{2\tih_t\sigma_n^2 \sigma_x^4}{\left(\tih_t^2\sigma_x^2+\sigma_n^2\right)^2} \right\}\nn\\
         &= \arg \max_{\vert\deh_t\vert\leq\eps} \left|\deh_t + \tih_t + \frac{l\bax}{w\baxs} - \frac{1}{w} + \frac{\tih_t\sigma_n^2 \sigma_x^4}{w^2\baxs\left(\tih_t^2\sigma_x^2+\sigma_n^2\right)^2}\right|.\label{eq:minmaxregretmax}
\end{align}
By applying the triangular inequality to the cost function in \eqref{eq:minmaxregretmax}, we get the following upper bound:
\begin{align*}
&\left|\deh_t + \tih_t + \dfrac{l\bax}{w\baxs} - \dfrac{1}{w} + \dfrac{\tih_t\sigma_n^2 \sigma_x^4}{w^2\baxs\left(\tih_t^2\sigma_x^2+\sigma_n^2\right)^2}\right|\\
&\leq \left|\deh_t \right| + \left|\tih_t + \dfrac{l\bax}{w\baxs} - \dfrac{1}{w} + \dfrac{\tih_t\sigma_n^2 \sigma_x^4}{w^2\baxs\left(\tih_t^2\sigma_x^2+\sigma_n^2\right)^2}\right| \\
&\leq \eps + \left|\tih_t + \dfrac{l\bax}{w\baxs} - \dfrac{1}{w} + \dfrac{\tih_t\sigma_n^2 \sigma_x^4}{w^2\baxs\left(\tih_t^2\sigma_x^2+\sigma_n^2\right)^2}\right|,
\end{align*}
where the upper bound is achieved at $\deh_t^* = \eps \mathrm{sgn}\left(\tih_t + \frac{l\bax}{w\baxs} - \frac{1}{w} + \frac{\tih_t\sigma_n^2 \sigma_x^4}{w^2\baxs\left(\tih_t^2\sigma_x^2+\sigma_n^2\right)^2}\right)$. Hence it follows that
\begin{align}
&\deh_t^*\nn\\
&=\arg \max_{\vert\deh_t\vert\leq\eps}\left\{ E\left[ \left(x_t - w\left((\tih_t+\deh_t) x_t + n_t\right) - l\right)^2 \right] - \frac{\sigma_n^2 \sigma_x^2}{\tih_t^2+\sigma_n^2} + \deh_t\frac{2\tih_t\sigma_n^2 \sigma_x^4}{\left(\tih_t^2\sigma_x^2+\sigma_n^2\right)^2} \right\}\nn\\
&= \left\{
     \begin{array}{ll}
       \eps & : \tih_t + \dfrac{l\bax}{w\baxs} - \dfrac{1}{w} + \dfrac{\tih_t\sigma_n^2 \sigma_x^4}{w^2\baxs\left(\tih_t^2\sigma_x^2+\sigma_n^2\right)^2}\geq0,\\
      -\eps & : \tih_t + \dfrac{l\bax}{w\baxs} - \dfrac{1}{w} + \dfrac{\tih_t\sigma_n^2 \sigma_x^4}{w^2\baxs\left(\tih_t^2\sigma_x^2+\sigma_n^2\right)^2}<0.\\
     \end{array}
     \right. \label{eq:minmaxregretdh}
\end{align}

We next solve the outer minimization problem as follows. If we write $\vu = \left[w,l\right]^T\in\mathbb{R}^2$ and define $\mathcal{M} = \left\{\vu=\left[w,l\right]^T\in\mathbb{R}^2\;\vert\;\tih_t + \frac{l\bax}{w\baxs} - \frac{1}{w} + \frac{\tih_t\sigma_n^2 \sigma_x^4}{w^2\baxs\left(\tih_t^2\sigma_x^2+\sigma_n^2\right)^2}\geq0\right\}$, then it follows that \\$\mathcal{N} \defi \left\{\vu=\left[w,l\right]^T\in\mathbb{R}^2\;\vert\;\tih_t + \frac{l\bax}{w\baxs} - \frac{1}{w} + \frac{\tih_t\sigma_n^2 \sigma_x^4}{w^2\baxs\left(\tih_t^2\sigma_x^2+\sigma_n^2\right)^2}<0\right\}=\mathbb{R}^2 \setminus\mathcal{M}$, i.e., $\mathcal{M} \cup \mathcal{N}=\mathbb{R}^2$ and $\mathcal{M} \cap \mathcal{N}= \varnothing$. Hence, the cost function in the outer minimization problem in \eqref{eq:minmaxregret3} is given by
\begin{align*}
&\max_{\vert\deh_t\vert\leq\eps} \left\{E\left[ \left(x_t - w\left((\tih_t+\deh_t) x_t + n_t\right) - l\right)^2 \right] - \frac{\sigma_n^2 \sigma_x^2}{\tih_t^2+\sigma_n^2} + \deh_t\frac{2\tih_t\sigma_n^2 \sigma_x^4}{\left(\tih_t^2\sigma_x^2+\sigma_n^2\right)^2}\right\} \nn\\
&=\left\{
     \begin{array}{ll}
       \left\{E\left[ \left(x_t - w\left((\tih_t+\eps) x_t + n_t\right) - l\right)^2 \right] - \frac{\sigma_n^2 \sigma_x^2}{\tih_t^2+\sigma_n^2} + \eps\frac{2\tih_t\sigma_n^2 \sigma_x^4}{\left(\tih_t^2\sigma_x^2+\sigma_n^2\right)^2}\right\} & : \left[w,l\right]^T\in\mathcal{M}\\
       \left\{E\left[ \left(x_t - w\left((\tih_t-\eps) x_t + n_t\right) - l\right)^2 \right] - \frac{\sigma_n^2 \sigma_x^2}{\tih_t^2+\sigma_n^2} - \eps\frac{2\tih_t\sigma_n^2 \sigma_x^4}{\left(\tih_t^2\sigma_x^2+\sigma_n^2\right)^2}\right\} & : \left[w,l\right]^T\in\mathcal{N}.\\
     \end{array}
     \right.
\end{align*}
We first substitute $\deh_t = \eps$ and find the corresponding $\{w,l\}$ pair
that minimizes the objective function in \eqref{eq:minmaxregret3} to check whether $\left[w,l\right]\in\mathcal{M}$.
We next substitute $\deh_t = -\eps$ and find the corresponding $\{w,l\}$ to check whether
$\left[w,l\right]\in\mathcal{N}$. Based on these criteria, we obtain the corresponding equalizer
impulse response and the bias term pair $\{w_{3,t},l_{3,t}\}$.

We first substitute $\deh_t = \eps$ in the cost function in \eqref{eq:minmaxregret3} to get the following
minimization problem:
\begin{align}
&\{w_1^*,l_1^*\} =  \arg \min_{w,l}\nn\\
&\left\{\baxs + w^2 \left(\tih_t+\eps\right)^2\baxs + w^2\sigma_n^2 + l^2 - 2w\left(\tih_t+\eps\right)\baxs -2\bax l - 2\bax w \left(\tih_t+\eps\right)l\nn\right.\\
&\left. \vphantom{}- \frac{\sigma_n^2 \sigma_x^2}{\tih_t^2+\sigma_n^2} + \eps\frac{2\tih_t\sigma_n^2 \sigma_x^4}{\left(\tih_t^2\sigma_x^2+\sigma_n^2\right)^2}\right\}. \label{eq:minmaxregretconvex1}
\end{align}
Since the cost function in \eqref{eq:minmaxregretconvex1} is a convex function of $w$ and $l$, we get that
\begin{align*}
w_1^*=\dfrac{\left(\tih_t +  \eps\right)\sigma_x^2 }{\left(\tih_t + \eps\right)^2 \sigma_x^2 +  \sigma_n^2},\;\;
l_1^*=\dfrac{\bax \sigma_n^2}{\left(\tih_t + \eps\right)^2\sigma_x^2 + \sigma_n^2}.
\end{align*}

We observe that $[w_1^*,l_1^*]\in\mathcal{M}$ if and only if
\begin{align*}
f\defi-\eps-\dfrac{\bax^2 \sigma_n^2}{\left(\tih_t + \eps\right)^2\sigma_x^2+\sigma_n^2} - \dfrac{\sigma_n^2}{\left(\tih_t + \eps\right)\sigma_x^2}
+ \dfrac{\tih_t\sigma_n^2}{\left(\tih_t + \eps\right)^2}\left(\dfrac{\left(\tih_t + \eps\right)^2\sigma_x^2 + \sigma_n^2}{\tih_t^2\sigma_x^2+\sigma_n^2}\right)^2\geq0.
\end{align*}

We next substitute $\deh_t = -\eps$ in the cost function in \eqref{eq:minmaxregret3} to get the following
minimization problem:
\begin{align}
&\{w_2^*,l_2^*\} =  \arg \min_{w,l}\nn\\
&\left\{\baxs + w^2 \left(\tih_t-\eps\right)^2\baxs + w^2\sigma_n^2 + l^2 - 2w\left(\tih_t-\eps\right)\baxs -2\bax l - 2\bax w \left(\tih_t-\eps\right)l\nn\right.\\
&\left. \vphantom{}- \frac{\sigma_n^2 \sigma_x^2}{\tih_t^2+\sigma_n^2} - \eps\frac{2\tih_t\sigma_n^2 \sigma_x^4}{\left(\tih_t^2\sigma_x^2+\sigma_n^2\right)^2}\right\}. \label{eq:minmaxregretconvex1}
\end{align}
Since the cost function in \eqref{eq:minmaxregretconvex1} is a convex function of $w$ and $l$, we get that
\begin{align*}
w_2^*=\dfrac{\left(\tih_t -  \eps\right)\sigma_x^2 }{\left(\tih_t - \eps\right)^2 \sigma_x^2 +  \sigma_n^2},\;\;
l_2^*=\dfrac{\bax \sigma_n^2}{\left(\tih_t - \eps\right)^2\sigma_x^2 + \sigma_n^2}.
\end{align*}

Note that $[w_2^*,l_2^*]\in\mathcal{N}$ if and only if
\begin{align*}
g\defi\eps-\dfrac{\bax^2 \sigma_n^2}{\left(\tih_t -\eps\right)^2\sigma_x^2+\sigma_n^2} - \dfrac{\sigma_n^2}{\left(\tih_t - \eps\right)\sigma_x^2}
+ \dfrac{\tih_t\sigma_n^2}{\left(\tih_t - \eps\right)^2}\left(\dfrac{\left(\tih_t - \eps\right)^2\sigma_x^2 + \sigma_n^2}{\tih_t^2\sigma_x^2+\sigma_n^2}\right)^2\leq0.
\end{align*}

There are four cases depending on the values of $\tih_t$, $\eps$, $\bax$, $\baxs$, $\sigma_n^2$:
\begin{itemize}
\item Case 1: $f\geq0$ and $g\geq0$.\\
In this case, we have
\begin{align*}
w_{3,t} = \dfrac{\left(\tih_t+\eps\right)\sigma_x^2}{\left(\tih_t+\eps\right)^2\sigma_x^2+\sigma_n^2}
\end{align*}
and
\begin{align*}
l_{3,t}=\dfrac{\bax\sigma_n^2}{\left(\tih_t+\eps\right)^2\sigma_x^2+\sigma_n^2}
\end{align*}
since $\left[w_1^*,l_1^*\right]\in\mathcal{M}$ and $\left[w_2^*,l_2^*\right]\not\in\mathcal{N}$.

\item Case 2: $f\leq0$ and $g\leq0$.\\
In this case, we have
\begin{align*}
w_{3,t} = \dfrac{\left(\tih_t-\eps\right)\sigma_x^2}{\left(\tih_t-\eps\right)^2\sigma_x^2+\sigma_n^2}
\end{align*}
and
\begin{align*}
l_{3,t}=\dfrac{\bax\sigma_n^2}{\left(\tih_t-\eps\right)^2\sigma_x^2+\sigma_n^2}
\end{align*}
since $\left[w_1^*,l_1^*\right]\not\in\mathcal{M}$ and $\left[w_2^*,l_2^*\right]\in\mathcal{N}$.

\item Case 3: $f\geq0$ and $g\leq0$.\\
In this case, we have $\left[w_1^*,l_1^*\right]\in\mathcal{M}$ and $\left[w_2^*,l_2^*\right]\in\mathcal{N}$ so that
\begin{align*}
&\left[w_{3,t},l_{3,t}\right] = \nn\\
&\arg \min_{\left[w,l\right]\in\left\{\left[w_1^*,l_1^*\right],\left[w_2^*,l_2^*\right]\right\}}\\
&\left\{\max_{\vert\deh_t\vert\leq\eps}\left\{ E\left[ \left(x_t - w\left((\tih_t+\deh_t) x_t + n_t\right) - l\right)^2 \right] - \frac{\sigma_n^2 \sigma_x^2}{\tih_t^2+\sigma_n^2} + \deh_t\frac{2\tih_t\sigma_n^2 \sigma_x^4}{\left(\tih_t^2\sigma_x^2+\sigma_n^2\right)^2} \right\} \right\}.
\end{align*}
\item Case 4: $f\leq0$ and $g\geq0$.\\
In the last case, we have the optimum points on the curve
$\tih_t + \frac{l\bax}{w\baxs} - \frac{1}{w} + \frac{\tih_t\sigma_n^2 \sigma_x^4}{w^2\baxs\left(\tih_t^2\sigma_x^2+\sigma_n^2\right)^2}=0$. Therefore $\deh_t^* = 0$ and the corresponding coefficients are given as the solution to the following optimization problem:
\begin{align*}
&\left[w_{3,t},l_{3,t}\right] = \arg \min_{\left[w,l\right]} \left\{ E\left[ \left(x_t - w\left(\tih_t x_t + n_t\right) - l\right)^2 \right] - \frac{\sigma_n^2 \sigma_x^2}{\tih_t^2+\sigma_n^2}\right\}\\
&\text{subject to }\\
&\tih_t + \frac{l\bax}{w\baxs} - \frac{1}{w} + \frac{\tih_t\sigma_n^2 \sigma_x^4}{w^2\baxs\left(\tih_t^2\sigma_x^2+\sigma_n^2\right)^2}=0.
\end{align*}
\end{itemize}
Hence, the proof follows. $\blacksquare$

\section{Simulations} \label{sec:sim}
We provide numerical examples in different scenarios in order to illustrate the performances of the equalization methods.
We first illustrate the performances of the channel equalization methods for a given perturbation bound. We demonstrate
that the the minimax equalization method yields the best worst case MSE performance among all methods for these simulations since it
optimizes the worst case MSE with respect to the worst case channel impulse response. We next present the average MSE performance of each method
over different channel perturbations. We show that the minimax regret method has better average MSE performance than the minimax
and minimin equalization methods for these simulations.

In the first set of experiments, we randomly generate a transmitted signal $x_t$
with mean 0.01 and variance 1.
We also generate a Gaussian channel noise $n_t$ with zero mean and unity variance.
The channel estimates are constructed using $\tih_t=h_t+\deh_t$, where
$h_t=1.05$ and the perturbation $\deh_t$ is randomly generated from a zero mean and $\eps$ standard
deviation Gaussian distribution and truncated to give
$\vert \deh_t \vert\leq\eps$ with $\eps=0.03$ for each trial.
Here, we label the method in Theorem 1 as ``Minimax", the method in Theorem 2 as
``Minimin", and finally the method in Theorem 3 as ``Minimax regret".
For each method and for each random perturbation, we find the corresponding
equalizer parameters $w_t$ and $l_t$ to calculate the estimates of the transmitted signal $x_t$.
After we calculate the mean-square errors for each method and for all random
perturbations, we plot the corresponding sorted errors in ascending
order in Fig.~2 for 200 trials.
Since the minimax equalization method optimizes the worst case MSE with
respect to worst possible perturbation, it yields the smallest worst case MSE
among all methods for these simulations. However, the overall
performance of the minimax method is significantly inferior to the
minimin and the minimax regret methods due to its highly conservative
nature. Furthermore, we notice that the minimax regret method provides
better average performance compared to the minimax and the minimin
methods and superior worst case performance compared to the minimin method
for these simulations.

\begin{figure}
\begin{center}
\includegraphics*[scale=0.45]{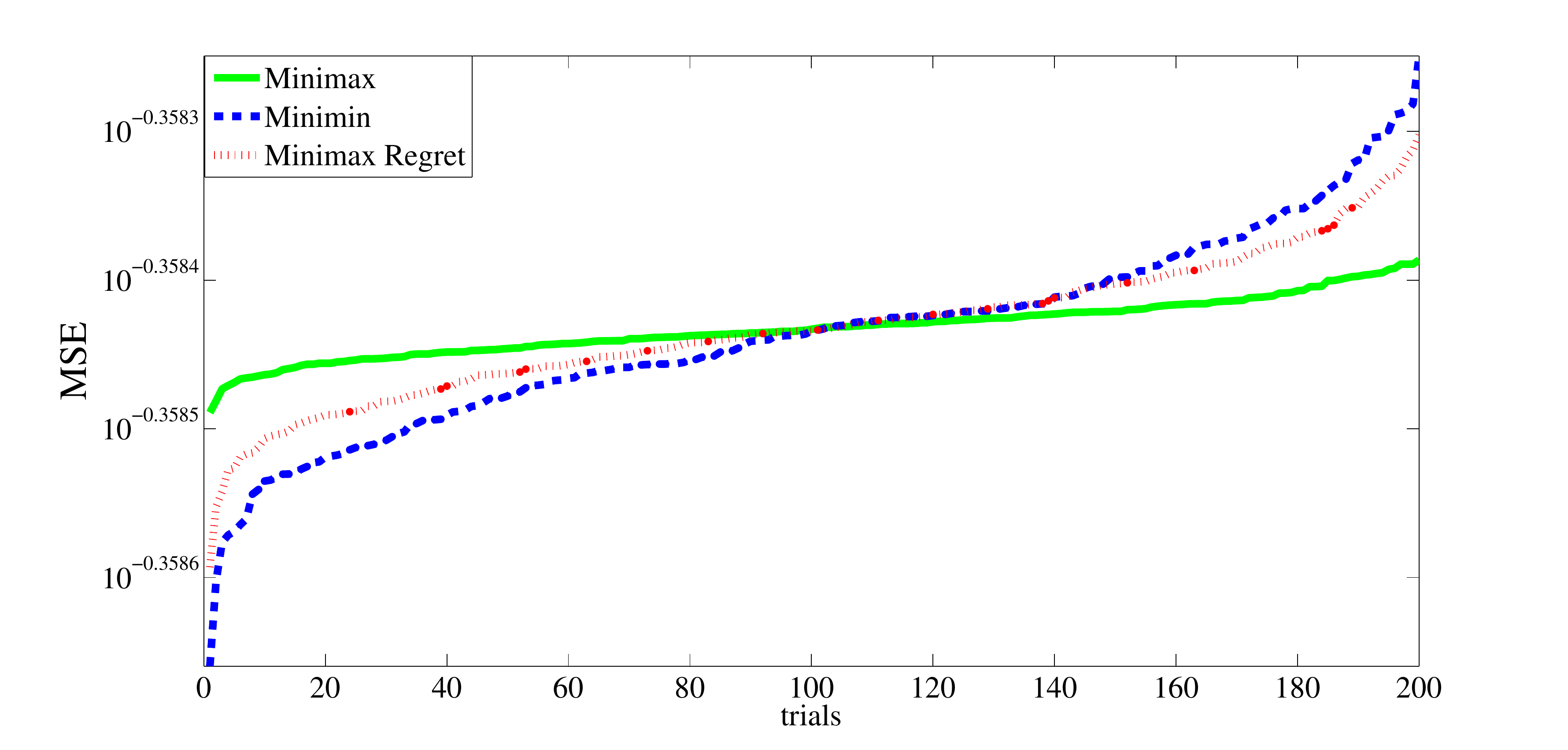}
\end{center}
\caption{Sorted MSEs for the minimax, minimin and minimax regret equalization methods over
200 trials when $\eps=0.3$.}
\label{fig:figure1}
\end{figure}

\begin{figure}
\begin{center}
\includegraphics*[scale=0.45]{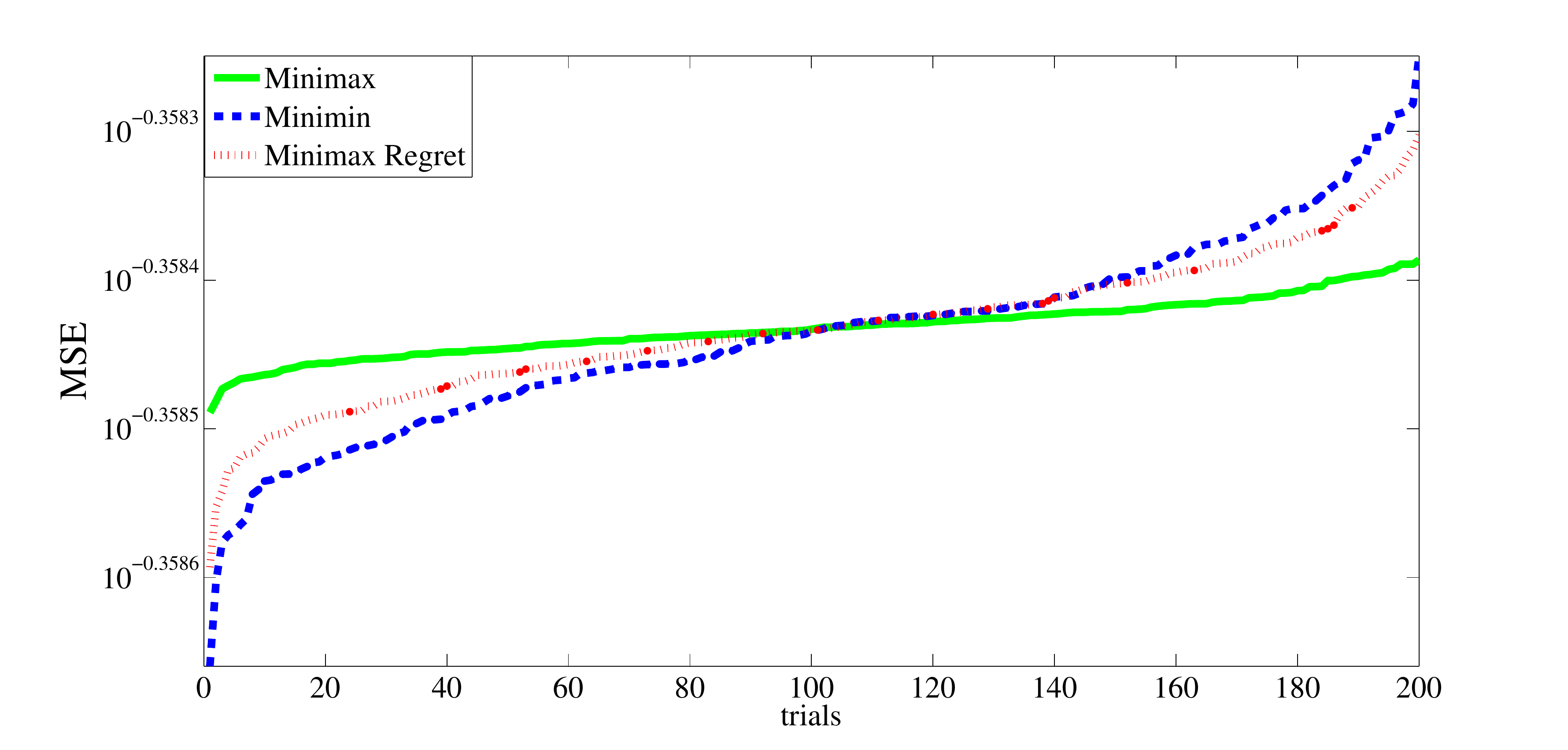}
\end{center}
\caption{Averaged MSEs for the minimax, minimin and minimax regret equalization methods over 200 trials when $\eps\in[0.1,0.3]$}
\label{fig:figure2}
\end{figure}

For the second experiment, we randomly generate $200$ random perturbations
$\deh_t$, where $\vert \deh_t \vert \leq \eps$ for different perturbation bounds
and compute the averaged MSEs over $200$ trials for the minimax, minimin
and the minimax regret methods. In this case, we randomly generate a
transmitted signal $x_t$ with zero and variance 1. The channel noise $n_t$ is generated from a Gaussian distribution
with zero mean and unity variance. Here, we construct the estimates of the channel
impulse response by $\tih_t=h_t+\deh_t$, where $h_t=1.05$ and the perturbation
$\deh_t$ is randomly generated from a zero mean and $\eps$ standard
deviation Gaussian distribution and truncated to give
$\vert \deh_t \vert\leq\eps$. In Fig.~3, we present the averaged MSEs
for each method where the perturbation bound varies, $\eps \in [0.1, 0.3]$.
We observe that the minimax regret method has the best average MSE performance
over different perturbation bounds compared to the minimax and the minimin equalization
methods.

In this section, we presented the performances of the minimax, minimin and minimax regret
channel equalization methods through simulations. We demonstrated that the minimax approach
leads to a better worst case MSE performances than the minimin and minimax regret approaches for these simulations.
We also presented the average MSE performances of the equalization methods over different channel perturbations
and showed that the minimax regret equalization method has the best average MSE performance among all methods
for these simulations.

\section{Conclusion} \label{sec:conc}
In this paper, we investigated the channel equalization problem for Rayleigh fading channels when
the channel impulse response is not accurately known.
We analyzed three robust methods to equalize Rayleigh fading channels that incorporate
the channel uncertainties into the problem formulation. We first studied the affine minimax channel equalization
framework that optimizes equalizer parameters to minimize the worst case MSE in the uncertainty region.
We next investigated  the affine minimin channel equalization method, which minimizes the MSE for the most favorable
channel impulse response in the perturbation bounds. Finally, we analyzed the affine minimax regret channel equalization framework, which
minimizes the worst case regret in the uncertainty region. We explicitly provide the equalizer coefficients
and the estimates of the desired signal for each method and for both zero mean
and nonzero signals. We illustrated the performances of these equalization
methods through simulations.

\bibliographystyle{elsarticle-num}
\bibliography{msaf_references}

\end{document}